\documentstyle[preprint,eqsecnum,aps]{revtex}

\def\half{\scriptstyle \frac {1} {2}}
\def\sech{{\rm sech}}
\begin{document} 
\draft
\preprint{TIFR/TH/97-20} 
\title{New Source Term for QGP Formation\\in the Background-Field
 Model}
\author{R. S. Bhalerao}
\address{Theoretical Physics Group\\
Tata Institute of Fundamental Research\\
Homi Bhabha Road, Colaba, Mumbai 400 005, India}
\author{V. Ravishankar}
\address{Department of Physics\\
Indian Institute of Technology\\
Kanpur 208 016, India}
\maketitle
\begin{abstract}
We consider pair production in a space-time-dependent background field
and derive a source term, i.e., production rate in the one-particle
phase space. Such a source term is required in
Boltzmann-equation-based models of quark-gluon plasma formation and
evolution. We compare the source term derived here with the one that
has been used in the literature so far. Significant differences are
observed.
\end{abstract}

\vfill


\noindent{E-mail: bhalerao@theory.tifr.res.in, ~~~~~~~~~~ 
Fax: 091 22 215 2110}

\noindent{E-mail: vravi@iitk.ernet.in, ~~~~~~~~~~~~~~~~~~~~
Fax: 091 0512 250260}

\newpage

\section{Introduction} 
In some models of ultrarelativistic nucleus-nucleus collisions, the
formation of quark-gluon plasma (QGP) is assumed to proceed via
creation of a strong colour-electric field in the region between the
two nuclei receding from each other after the collision, and the
subsequent decay of the field by parton-pair production according to
the well known nonperturbative formula given by Schwinger \cite{sch}.
We shall call these models \cite{km,gat,bial,ban,am,je}
background-field models. They make use of the framework of the
relativistic transport theory to study the formation and
preequilibrium evolution of QGP, with Schwinger's formula providing a
basis for derivation of the source term (i.e., production rate in the
one-particle phase space) in the Boltzmann transport equation.

In analogy with an ordinary parallel-plates capacitor, the above
models have also been called colour-plates models. The colour
field, however, need not owe its existence entirely to the colour
plates. A high-density many-body system is often conveniently
described in terms of a self-consistent mean field and residual
short-range interactions. For example, Blaizot and Iancu
\cite{blaizot} have considered gauge theories at high temperatures and
shown that long-wavelength excitations of QGP can be described as
collective oscillations of gauge and fermionic mean fields. They have
presented non-abelian solutions to the equations of motion, which
correspond to spatially uniform colour oscillations. Importance of the
mean field has also been stressed in the context of hot and dense
hadron gas models of relativistic heavy ion collisions; see for
example \cite{li}.  Whatever may be the origin of the colour
field, once it is there, it can produce pairs of partons, add to the
entropy of the system and affect its subsequent evolution.

A few remarks on the source term mentioned above are in order. The
exact one-loop nonperturbative result (Fig. 1a) that Schwinger derived
for pair production in a constant, uniform background electric field
($E$) is given by
\begin{equation} 
{dW \over d^3x~dt} = {\alpha E^2 \over \pi^2}
\sum_{n=1}^\infty {1 \over n^2} \exp \left( -{n \pi m^2 \over 
\vert eE \vert} \right), 
\end{equation}
where $\alpha = e^2/4\pi$ is the fine-structure constant and $m$ is
the particle mass. It is instructive to understand this result in
terms of a semiclassical tunneling across the mass gap \cite{cash}.
This result, strictly speaking, is applicable to the parton pair
production only in the so-called abelian approximation. Secondly,
Eq. (1.1) gives the probability per unit volume and per unit time,
that a pair is produced, while the transport equation requires the
source term in {\it one-particle phase space}. Finally, and more
importantly, unlike the case considered in \cite{sch} where the
constant electric field is maintained by external sources, the pairs
produced in the background-field models derive their energy and
momentum from the field, the field becomes time dependent and
Eq. (1.1) stops being applicable.

The resolution of these problems has so far involved rewriting the
right-hand side of Eq. (1.1) as an integral over $p_T^2$ or
$m_T^2 \equiv p_T^2 + m^2$, $~p_T$ being the parton momentum
transverse to the direction of the collision axis. The expression so
obtained has often been used in the QGP transport equation and it
reads \cite{cash}
\begin{equation}
{dW \over d^3x~dt~d^2p_T} = {1 \over 4\pi^3} \left\vert eE 
\ln \left[ 1 - \exp \left( -{\pi m_T^2 \over \vert eE \vert} 
\right)\right] \right\vert.
\end{equation}
While the above extension may be justified semiclassically, the next
assumption that the constant $E$ in Eq. (1.2), may be replaced by its
time-dependent counterpart $E(t)$, merits a critical examination.
Strictly speaking, this replacement is invalid since as soon as the
field acquires time dependence, it can directly excite negative-energy
particles to levels above the mass gap, by a perturbative mechanism,
without recourse to any tunneling or barrier penetration mechanism.
The above replacement may be justified as a kind of a local-field
approximation provided the time-dependence of $E(t)$ is so weak that
it only excites frequencies with magnitudes negligible compared to the
mass gap. Equivalently, the field needs to be constant on the scale of
the Compton wavelength $m^{-1}$. Fields that are constant over such
large time intervals are unlikely to be produced in heavy-ion
collisions. Moreover, a justification for the local-field
approximation, which could be looked for {\it a posteriori} is not
forthcoming from the explicit numerical results presented in
\cite{ban}.

The purpose of this paper is to take the time dependence of the
electric field into account, as properly as a semiclassical formalism
would allow, and present a new source term based on Fig. 1b, for use
in the QGP transport equation, thereby avoiding the unphysical aspect
that we have pointed out in the current formulation. The mechanism
depicted in Fig. 1b was used by Martin et al. \cite{MVC} to
calculate the total energy deposited during a certain time interval,
under certain assumptions. They did not derive a source term.

\section{New Source Term}

For fermion pair production in the presence of a {\it
space-time-dependent} background field the amplitude to the lowest
order (Fig. 1b) is given by
\[
-i e~ \bar u(p_1)~ A\!\!\!/(K)~ v(p_2), 
\]
and the probability for pair production is
\begin{equation}
W = {\alpha \over \pi} \int d^4K~ \delta^{(4)}(K-p_1-p_2) \int
{d^3p_1 \over 2E_1}~ {d^3p_2 \over 2E_2}~ T,
\end{equation}
where
\[
T \equiv {\rm tr}~ [({\not p}_1 + m)~A\!\!\!/(K)~ ({\not p}_2 - m) 
~A\!\!\!/(-K)].
\]
Since our aim is to obtain the production rate in the one-particle
phase space, for insertion in the Boltzmann equation for a
one-particle distribution function $f(x,p)$, we choose to perform the
$K$-integration first. Next, in order to evaluate $T$, we introduce
\begin{equation}
P \equiv p_1+p_2 ~{\rm and}~ p \equiv p_1 - p_2,
\end{equation}
and use $p_1 \cdot p_2 + m^2 = P^2/2$. This gives 
\begin{eqnarray}
T &=& (P+p) \cdot A(P)~ (P-p) \cdot A(-P) - 2 P^2 A(P) \cdot A(-P)+
(P+p) \cdot A(-P)~(P-p) \cdot A(P) \nonumber \\
&=& 2~P \cdot A(P)P \cdot A(-P)-2~P^2A(P) \cdot A(-P)-2~p \cdot A(P)
~p\cdot A(-P).
\end{eqnarray}
Now recall
\begin{eqnarray}
- {1 \over 2}  F_{\mu \nu}(P)F^{\mu \nu}(-P) &=& P \cdot A(P)~P 
\cdot A(-P) -P^2A(P) \cdot A(-P) \nonumber \\
&=& |{\bf E}(P)|^2 - |{\bf B}(P)|^2.
\end{eqnarray}
To be able to compare our results with those based on Eq. (1.2), we
ignore the magnetic field ${\bf B}(P)$. Note also that pair production
is an electric effect \cite{sch}. Equations (2.3) and (2.4) give us
\begin{equation}
T=2~|{\bf E}(P)|^2 -2~p \cdot A(P)~p \cdot A(-P).
\end{equation}

Note that the remaining term $p \cdot A(P)~p \cdot A(-P)$ in 
Eq. (2.5) is also invariant under the gauge transformation
\[
A_\mu(P) \rightarrow A_\mu(P) - i~ P_\mu~\chi(P),
\]
because $p \cdot P=0$. We shall now rewrite this term also in terms of
${\bf E}(P)$. To that end, we choose the following gauge:
\begin{eqnarray}
A^1 &=& 0 =A^2, \nonumber \\
A^\alpha(x) &=& \epsilon^{\alpha \beta}~ \partial _\beta a(x),~~~~~
(\alpha, \beta = 0 ~{\rm or}~ 3)
\end{eqnarray}
where $a(x) = a(t,z)$
is an arbitrary real function vanishing at $t$ and $z$ $=
\pm \infty$, and $\epsilon^{\alpha \beta}$ is the totally
antisymmetric Levi-Civita tensor in $(1+1)$ dimensions. 
(This gauge choice is natural in the colour-plates model where $\bf E$
is taken along the collision or $z$-axis.)
Equation (2.6)
ensures the gauge condition $\partial _\mu A^\mu(x) = 0$. We have
\begin{eqnarray}
A^\alpha (P) &=& (2 \pi)^{-2} \int d^4x~ \exp (-iP \cdot x)~ 
A^\alpha (x) \nonumber \\
&=& i~ \epsilon ^{\alpha \beta}~ P_\beta~ (2\pi)^{-2} \int
d^4x ~\exp (-iP \cdot x)~ a(x)
\nonumber \\
&=& i~ \epsilon ^{\alpha \beta}~ P_\beta~ a(P)
\nonumber \\
&\equiv& i~ \tilde {P}^\alpha~ a(P),
\end{eqnarray}
where $\tilde {P}$ is the dual of $P$.

Keeping in mind the boost-invariant central-rapidity region expected
to be formed in ultrarelativistic heavy-ion collisions, we assume that
the electric field ${\bf E}(x)$ depends only on the boost-invariant
variable $\tau \equiv \sqrt{t^2-z^2}$. Hence
\begin{eqnarray}
{\bf E}(P) &=& (2\pi)^{-2} \int d^4x~ \exp(-i P \cdot x)~ {\bf E}(x) 
\nonumber \\
&=& \delta^{(2)}({\bf P}_T)\int dt ~dz~ \exp [-i(P_0t-P_zz)]~
{\bf E}(\tau).
\end{eqnarray}
It follows that ${\bf p}_{1T}=-{\bf p}_{2T}$ and hence the two
particles have identical transverse mass, say $m_T$. The relevant
components of their 4-momenta are
\begin{mathletters}
\label{generallabel}
\begin{equation}
p_1^\alpha=m_T(\cosh y_1,~\sinh y_1), ~~~~p_2^\alpha=m_T(\cosh y_2,
~\sinh y_2),
\label{mlett:1}
\end{equation}
such that
\begin{equation}
\tilde p_1^\alpha=-m_T(\sinh y_1,~\cosh y_1),~~~~ \tilde p_2^\alpha
=-m_T(\sinh y_2,~\cosh y_2),
\label{mlett:2}
\end{equation}
\end{mathletters}
where $y_{1,2}$ are their rapidities. For a future reference we note
the following results
\begin{eqnarray}
p_1 \cdot \tilde p_1 &=& 0 =p_2 \cdot \tilde p_2, ~~~~\tilde p_1 \cdot 
p_2= -p_1 \cdot \tilde p_2,\\
P \cdot \tilde P &=& 0, ~~\tilde P^2 = -P^2, \\
P^2 &=& (p_1+p_2)^2=2~m_T^2~[1+\cosh(y_1-y_2)].
\end{eqnarray}

Now in order to rewrite $p \cdot A(P)~p \cdot A(-P)$ in terms of
${\bf E}(P)$, we use Eqs. (2.7), (2.2), (2.10) and (2.9) to get
\begin{equation}
p \cdot A(P)~p \cdot A(-P) = 4~m_T^4~ \sinh^2(y_1-y_2)~a(P)a(-P).
\end{equation}
We also note from Eqs. (2.4), (2.7) and (2.11) that
\begin{equation}
|{\bf E}(P)|^2 = (P^2)^2~a(P)a(-P).
\end{equation}
Finally, eliminating $a(P)a(-P)$ between Eqs. (2.13) and (2.14) we get
the desired expression
\begin{equation}
p \cdot A(P)~p \cdot A(-P) = 4~m_T^4~ \sinh^2(y_1-y_2)~|{\bf E}(P)|^2
~/~(P^2)^2.
\end{equation}
The occurrence of $(P^2)^2$ in the denominator is noteworthy. It
indicates a highly nonlocal nature of this term in
the configuration space; the production rate depends
not only on the instantaneous strength of the field but also on its
derivatives. {\it This feature is absent in Eq. (1.2).} Substituting
$P^2$ from Eq. (2.12) in Eq. (2.15) and then substituting the
resulting expression in Eq. (2.5) we get
\[
T=2~|{\bf E}(P)|^2 ~\sech^2 \left({y_1-y_2 \over 2}\right).
\]
Thus the probability $W$ in Eq. (2.1) becomes
\begin{equation}
W={\alpha\over 2\pi}\int {d^3p_1\over E_1}~{d^3p_2\over E_2}~
|{\bf E}(P)|^2~
\sech^2 \left({y_1-y_2 \over 2}\right).
\end{equation}

In order to evaluate ${\bf E}(P)$ occurring in Eq. (2.16) we define (for
$t>0, ~t^2>z^2$)
\begin{eqnarray}
\tau^2&=&t^2-z^2, ~~\eta = \tanh^{-1}(z/t), \nonumber \\
M^2&=&P_0^2-P_z^2,~~ Y= \tanh^{-1}(P_z/P_0), \nonumber
\end{eqnarray}
and substitute
\begin{eqnarray}
dt~dz&=&\tau ~d \tau ~d \eta, \nonumber \\
P_0t-P_zz&=&M ~\tau~ \cosh(Y-\eta) \equiv M~ \tau~ \cosh \theta, 
\nonumber
\end{eqnarray}
in Eq. (2.8) to get
\begin{equation}
E(P)=\delta^{(2)}({\bf P}_T) \int_{-\infty}^\infty d \theta 
\int_0^\infty d\tau ~\tau~ \exp(-iM \tau \cosh \theta)~E(\tau).
\end{equation}
We shall assume for the sake of definiteness and simplicity that
$E(\tau)$ depends on only one time scale ($\tau_0$) and is of the form
\begin{equation}
E(\tau)=E_0~\exp(-\tau/\tau_0),~~~\tau_0>0.
\end{equation}
Self-consistent numerical calculations reported in Ref. \cite{ban}
predict $\tau$-dependence which is consistent with Eq. (2.18).
Substituting $E(\tau)$ given in Eq. (2.18) in Eq. (2.17) and
performing the $\tau$-integration by parts, we get
\[
E(P)=E_0 ~\delta^{(2)}({\bf P}_T)~\tau_0^2 \int_{-\infty}^\infty
d \theta~(1+iM \tau_0 \cosh \theta)^{-2}.
\]
The $\theta$-integration too can be done analytically \cite{grad}
which gives
\begin{equation}
E(P)=E_0 ~\delta^{(2)}({\bf P}_T)~ \tau_0^2~2~I(M \tau_0)~/~
(1+M^2 \tau_0^2),
\end{equation}
where
\begin{equation}
I(x) \equiv -1+{1\over \sqrt{1+x^2}}
\ln{1+ix+\sqrt{1+x^2} \over 1+ix-\sqrt{1+x^2}}.
\end{equation}

We continue with our derivation of $W$ by substituting $E(P)$ from
Eq. (2.19) in Eq. (2.16). This entails the calculation of
$[\delta^{(2)}({\bf P}_T)]^2$ which we perform as follows:
\begin{eqnarray}
[\delta^{(2)}({\bf P}_T)]^2 &=& \delta^{(2)}({\bf P}_T)\int
\exp(i{\bf P}_T \cdot {\bf x}_T)~ d^2x_T~/~(2\pi)^2 \nonumber \\
&\rightarrow& \delta^{(2)}({\bf P}_T)~d^2x_T~/~(2\pi)^2.
\end{eqnarray}
This is a standard procedure and it allows us to write the probability
per unit transverse area. Next, recall the occurrence of $(P^2)^2$ in
Eq. (2.15) and the comment made thereafter. In order to recover the
dependence of $W$ on the instantaneous strength of the field
$|E(\tau)|^2$ and to obtain a time-dependent 
rate, we shall employ the integral representation:
\begin{equation}
{\half}  \tau_0=\int_0^\infty\exp(-2\tau/\tau_0)~d\tau.
\end{equation}
Equations (2.19)-(2.22) together with Eq. (2.16) yield
\begin{eqnarray}
dW=\alpha\left({\tau_0\over\pi}\right)^3&&d^2x_T~d\tau~|E(\tau)|^2
\int{d^3p_1 \over E_1}{d^3p_2 \over E_2}~ \delta^{(2)}({\bf P}_T)
\nonumber\\
&{\times}& |I(M\tau_0)|^2~\sech^2 \left({y_1-y_2\over 2}\right)
(1+M^2\tau_0^2)^{-2}. \nonumber
\end{eqnarray}
Note that
\[
{d^3p_1 \over E_1}{d^3p_2 \over E_2}=d^2p_{1T}~dy_1~d^2p_{2T}~dy_2.
\]
The integration over ${\bf p}_{2T}$ can be done trivially with the
help of the delta function $\delta^{(2)}({\bf P}_T={\bf p}_{1T} +{\bf
p}_{2T}).$ In order to obtain the
production rate in the one-particle phase space we follow the
literature and assume as in Ref. \cite{km} that if a particle is
produced at $z$ and $t$, it must appear with the longitudinal velocity
$p_z/E=z/t$. (This is reminiscent of the Thomas-Fermi approximation in
condensed-matter physics.) It allows us to identify the spatial
and momentum rapidities and we can write $d \tau ~dy_1=dt~dz/\tau.$

Combining these results we finally get the new source term:
\begin{equation}
{dW \over d^3x~dt~d^2p_T}=\alpha\left(\tau_0 \over \pi\right)^3
{|E(\tau)|^2 \over \tau}\int_{-\infty}^\infty|I(M\tau_0)|^2
~\sech^2 \left({y_1-y_2 \over 2}\right)(1+M^2\tau_0^2)^{-2}d(y_1-y_2),
\end{equation}
where the electric field $E(\tau)$ and the function $I(M\tau_0)$ are
defined in Eqs. (2.18) and (2.20), respectively, and from Eq. (2.12)
the invariant mass of the pair, $M$, is given by
\[
M=2~m_T~ \cosh {\half} (y_1-y_2).
\]

\section{Results}

We now compare the old and the new source terms given in Eqs. (1.2)
and (2.23), respectively. Both are invariant under Lorentz boosts in
the longitudinal or $z$ direction. In the limit $E \rightarrow 0$,
(1.2) and all of its derivatives vanish, while (2.23) behaves like
$\sim E^2$. In the large $E$ limit, (1.2) behaves like $\sim E~
\ln(E)$, while (2.23) behaves like $\sim E^2$. Thus the ratio of the
new and the old source terms increases with the strength of the
electric field. The transverse-momentum ($p_T$) dependence arises
naturally in the derivation of Eq. (2.23). The longitudinal-momentum
dependence too can be recovered, if necessary.

We also studied these two terms numerically as a function of the
proper time $\tau$ and the transverse momentum $p_T$, for various
values of the input parameters, namely the particle mass $m$, the
parameter $\tau_0$ appearing in Eq. (2.18), the fine-structure
constant $\alpha$, and the initial strength of the electric field
$E_0$. Some representative results are shown in Figs. (2)-(4), where
we have taken $m = 5$ MeV, $\tau_0 = 0.1$ fm, $\alpha = 0.2$, and
${\half} E_0^2 = 10$ GeV/fm$^3$. Significant differences are evident.

In view of the above comments and the numerical results presented
here, the two source terms, Eqs. (1.2) and (2.23), when substituted in
the Boltzmann equation are expected to contribute differently to the
formation of the quark-gluon plasma. It is straightforward to extend
Eq. (2.23) to the case of {\it coloured} quark and antiquark
production, by introducing an appropriate colour factor. A similar
source term for gluon production in a space-time-dependent colour
field would also be required; that work is in progress \cite{NR1}. It
would be interesting to substitute these new source terms in a {\it
nonabelian} transport equation \cite{NR2}, and study formation of
quark-gluon plasma. We plan to undertake this work in the future.

\acknowledgments
We thank M. Guchait and N. Parua for drawing the diagrams in Fig. 1.

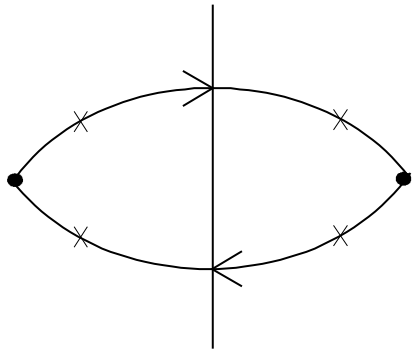
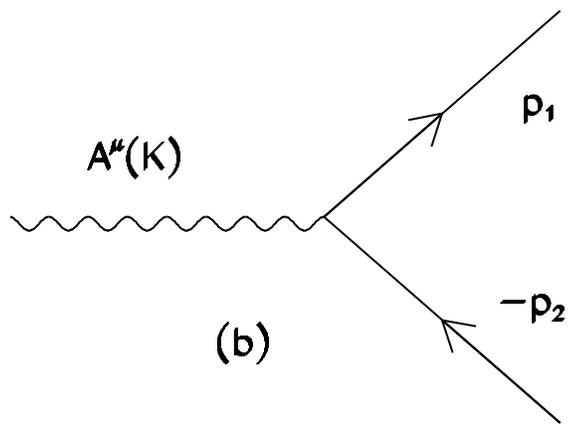
\begin{figure} 
\caption
{Schematic diagrams for pair production in (a) time-independent and
(b) time-dependent electric fields. In (a), crosses indicate
interactions (to all orders) with the vacuum and the vertical line
represents the ``cut''. (b) represents pair production in the
lowest-order perturbation theory.}
\label{1}
\end{figure}

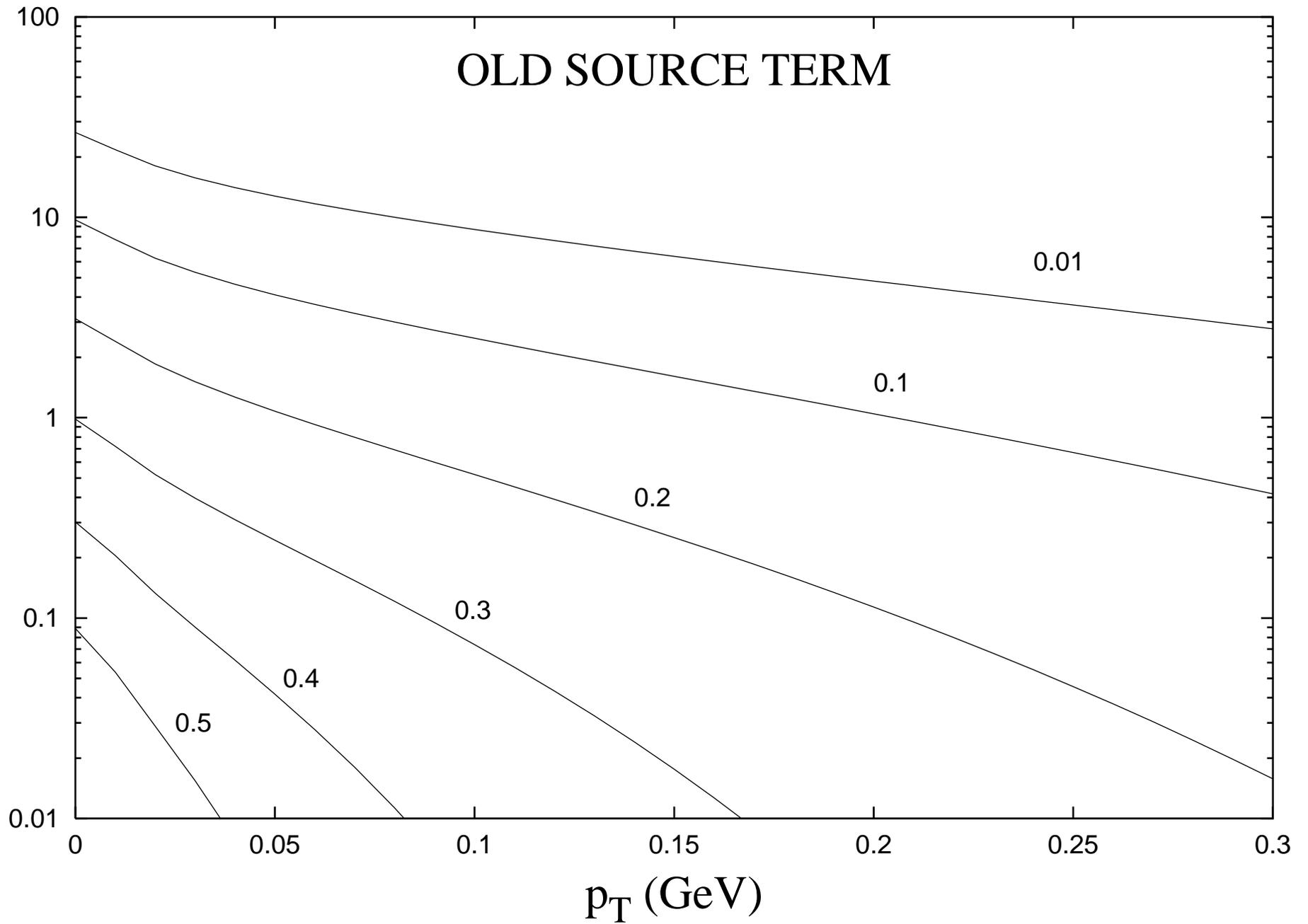
\begin{figure} 
\caption
{The old source term, Eq. (1.2), in units of fm$^{-4}$ GeV$^{-2}$, as
a function of the transverse momentum $p_T$ (GeV), for various values 
of the proper time $\tau$. The curves are labelled by $\tau$ (fm).
Input parameters are as in the text.}
\label{2}
\end{figure}

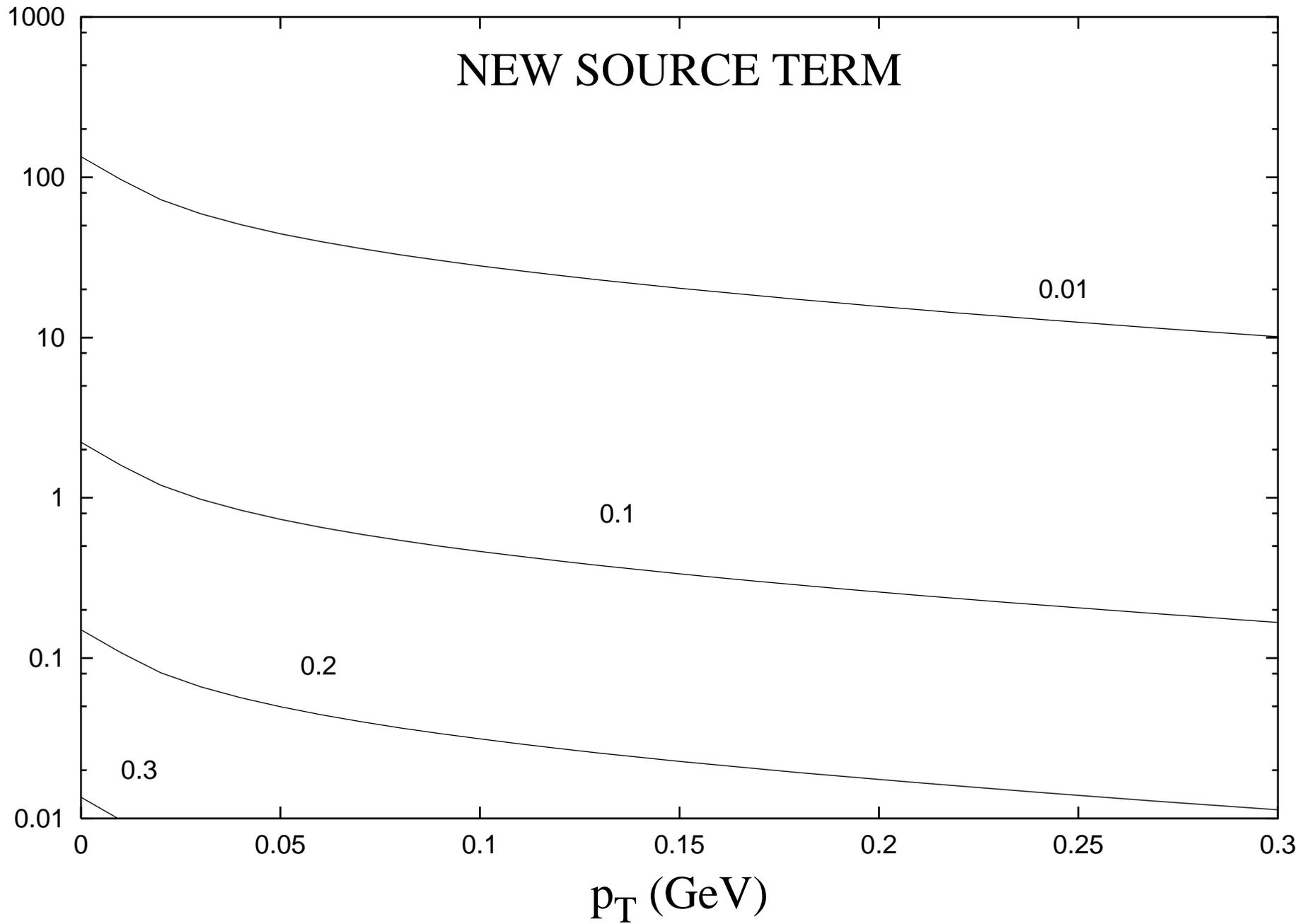
\begin{figure} 
\caption
{Same as Fig. 2, for the new source term, Eq. (2.23). Note the 
difference in the scale.}
\label{3}
\end{figure}

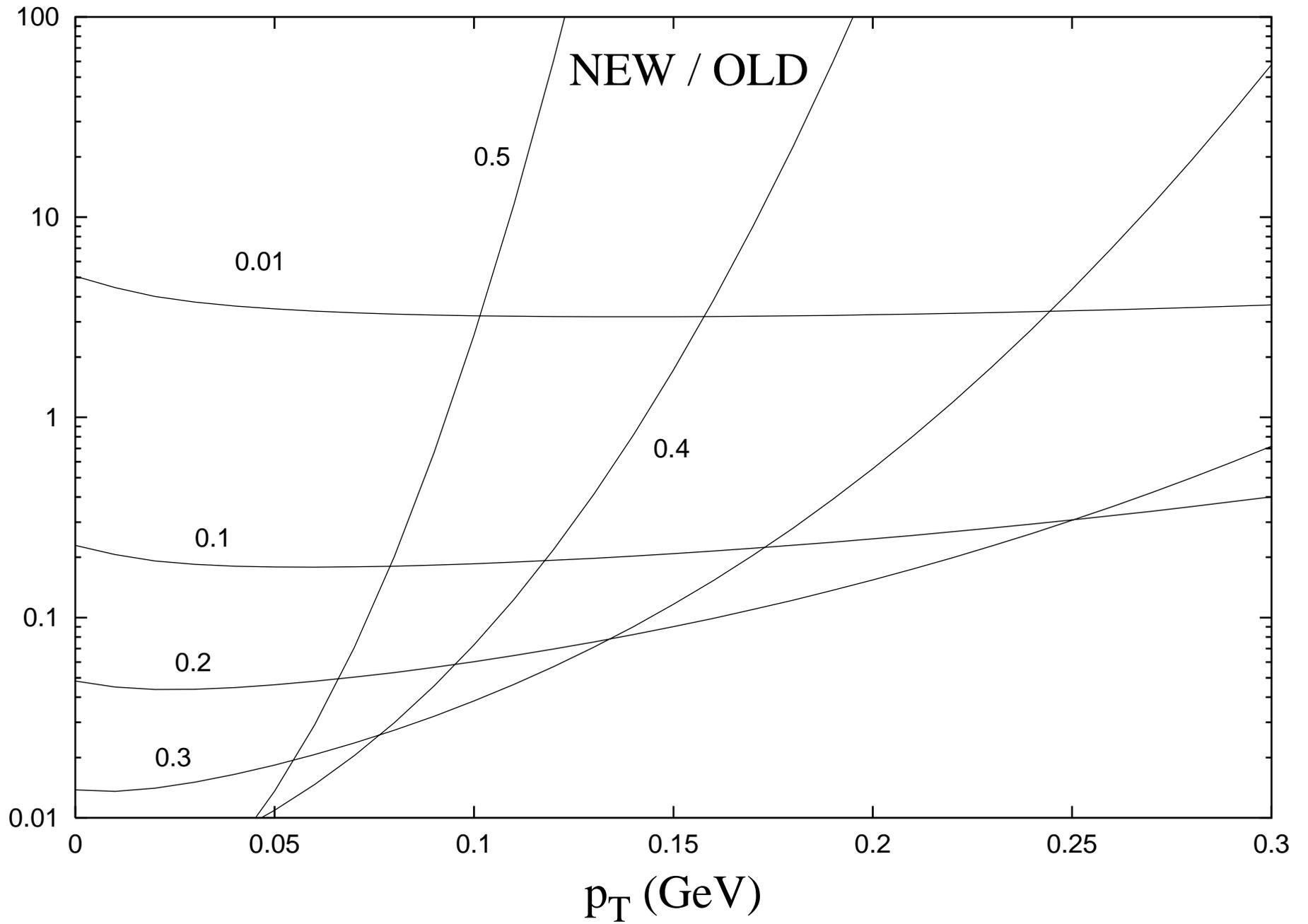
\begin{figure} 
\caption
{The ratio of the new and the old source terms. The curves are labelled
by $\tau$  (fm).}
\label{4}
\end{figure}

\end{document}